\documentclass[preprint,prb]{revtex4}
\usepackage{graphicx}
\usepackage{dcolumn}
\usepackage{bm}

\begin{document}
\begin{titlepage}
\title{Energetics and magnetism of Co-doped GaN(0001) surfaces: A first-principles study}

\author{Zhenzhen Qin$^1, ^2$, Zhihua Xiong$^1$\footnote{E-mail: xiong\_zhihua@126.com}, Guangzhao Qin$^3$, Lanli Chen$^1$}\address{$^1$Key Laboratory for Optoelectronics and Communication of Jiangxi Province, Jiangxi Science \& Technology Normal University \\Nanchang, 330018, People's Republic of China\\$^2$College of Electronic Information and Optical Engineering, Nankai University, Tianjin, 300071, Peoples Republic of China \\
$^3$College of Materials Science and Opto-Electronic Technology, University of Chinese Academy of Sciences, Beijing, 101408, People's Republic of China}


\begin{abstract}
A comprehensive first-principles study of the energetics, electronic and magnetic properties of Co-doped GaN(0001) thin films are presented and the effect of surface structure on the magnetic coupling between Co atoms is demonstrated. It is found that Co atoms prefer to
substitute the surface Ga sites in different growth conditions.  In particular, a CoN/GaN interface structure with Co atoms replacing the first Ga layer is preferred under N-rich and moderately Ga-rich conditions, while CoGa$_x$/GaN interface is found to be energetically stable under extremely Ga-rich conditions. It's worth noted that the antiferromagnetic coupling between Co atoms is favorable in clean GaN(0001) surface, but the existence of FM would be expected to occur as Co concentration increased in Ga-bilayer GaN(0001) surface. Our study provides the theoretical understanding for experimental research on Co-doped GaN films and might promise the Co:GaN system potential applications in spin injection devices.

\end{abstract}

\maketitle
\draft
\vspace{2mm}
\end{titlepage}
\maketitle

\section{Introduction}
Diluted magnetic semiconductors (DMS) have much attention due to their potential applications as spintronic devices\cite{H. Ohno, Shifei Qi, Alexander A. Khajetoorians, Xu Zuo}.  In recent years, Co:GaN system is of particularly importance due to its potential applications in magneto electronics.  Experimental and theoretical studies have been reported\cite{G. H. Zhong, H D Li, S. Dhara, W.Kim, V. Baranwal, S. Munawar Basha} to investigate magnetic properties in Co:GaN films.  A strong ferromagnetic(FM) ordering with Tc extending $\sim$250K is found by Dhara et.al \cite{S. Dhara} in Co-doped n-GaN by metal-organic chemical vapor deposited(MOCVD).  To explore the magnetism origination, Kim\cite{W.Kim} supposed that the magnetism originates from the CoGa$_x$ or Co clusters in Co-doped GaN films. Based on the effect of doped concentration, Munawar Basha et. al\cite{S. Munawar Basha} concluded that only lower percentage of Cobalt doped GaN is suitable for spintronic applications and supposed that the decrease in magnetic moment is attributed to the formation of secondary phase or Cobalt cluster, which contributes to the antiferromagnetic(AFM) states. However, the microscopic nature of magnetism in the Co-doped GaN films has remained unclear.

Moreover, there are many meaningful studies on magnetic characteristics in Co-doped ZnO surfaces \cite{Yu-feng Tian, W. G. Xie, A. Ney, Numan Akdogan}, which motivate us to explore the magnetism in the Co-doped GaN surfaces.
In addition, the structural properties of GaN surfaces depend sensitively on the orientation of the surface termination and reconstructions\cite{G. Mula,
PhysRevB.69.035325}. The GaN(0001) surface has been demonstrated having a better
surface morphology, and an incommensurate laterally contracted Ga
bilayer structure\cite{J. E. Northrup} can exist in extremely Ga-rich conditions, which
are typical of molecular-beam epitaxy(MBE) growth. It is believed that the surface structure and morphology are more sensitive to magnetic exchange interactions. Therefore, it is essential to explore the magnetism
of Co-doped GaN(0001) under different growth conditions, which is important for the better understanding of the interfacial
implications of heterojunction thin films in further experiments.

In this paper, energetics and magnetic properties of Co-doped clean and
Ga-bilayer GaN(0001) surfaces are systematically studied by first-principles
calculations.
It is found that Co atoms show a strong surface sites preference, while different
interface structures would be formed in the two surfaces above.
The magnetic calculations show that the preferred state of Co-doped GaN(0001) thin film is antiferromagnetic under N-rich conditions, but becomes ferromagnetic with the increase in Co concentration under extremely Ga-rich conditions. Finally, electronic structure calculations for the AFM and FM configurations are performed to explain the magnetic mechanism in two different GaN(0001) surfaces respectively.

\section{COMPUTATIONAL DETAILS}
\begin{figure}
    \begin{minipage}{\linewidth}
        \includegraphics[width=\textwidth]{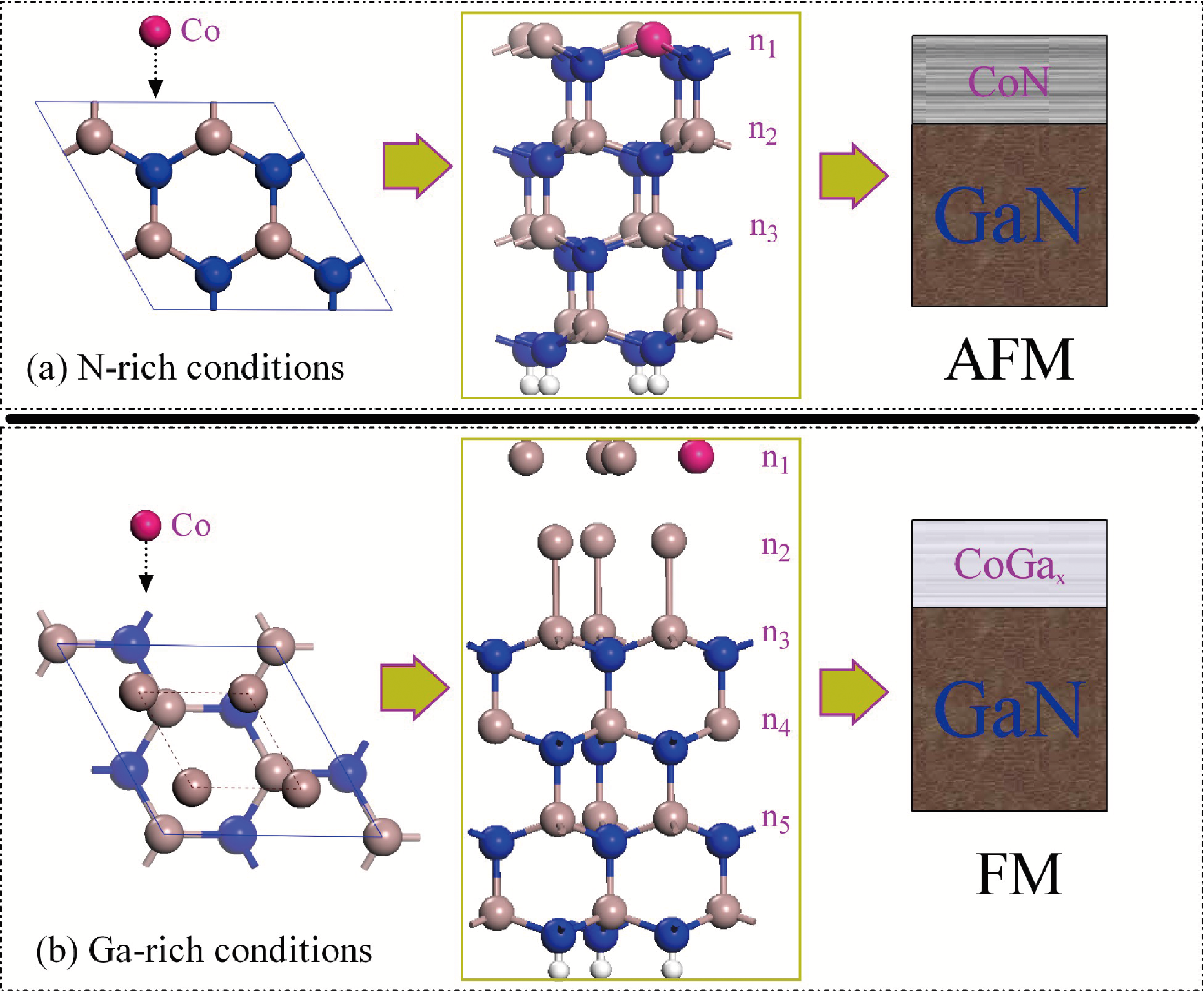}
    \end{minipage}
\caption{\label{Fig1}
(Color online)
(a) (top-view and side-view) Models of Co-doped clean GaN(0001)
surface, corresponding to the N-rich conditions or moderately Ga-rich conditions.
(b) (top-view and side-view) Models of Co-doped Ga-bilayer GaN(0001) surface, corresponding to
the extremely Ga-rich conditions. The vacuum regions are not shown here.}
\end{figure}

Our calculations are performed using the projected augmented
wave (PAW)\cite{P.E.Blochl} method as implemented in the Vienna \emph{ab-initio}
simulation package (\texttt{\textsc{vasp}})\cite{G.K}.
Exchange-correlation energy functional is treated in the Perdew-Burke-Ernzerhof of generalized gradient
approximation(GGA-PBE)\cite{J. P. Perdew}.
The valence electron configurations for Co, Ga and N are considered as
$3d^{7}4s^{2}$, $3d^{10}4s^{2}4p^{1}$ and $2s^{2}2p^{3}$, respectively.
The wave functions are expanded in plane wave basis with a kinetic energy cut
off of 450 eV.
Then, to study the behavior of Co atoms in GaN(0001) surface
under different growth conditions, the clean GaN(0001)-$2\times 2$ surface and the Ga-bilayer GaN(0001)-$\sqrt{3}\times \sqrt{3}$ surface shown in  Fig.~\ref{Fig1} are modeled to
simulate the slab films under N-rich and extremely Ga-rich conditions respectively. Here we use the optimized lattice constants of wurtzite GaN cell with a=3.216 $\textrm{\AA}$ and c=5.239 $\textrm{\AA}$ as listed in Table I of Ref.\cite{Zhenzhen Qin}.
By testing the optimum for slab thickness, we find that the deepest cobalt atoms do not exhibit a significant spurious interaction with the bottom surface for four GaN bilayer system.
Thus, in this work, we adopt the surface configuration including four GaN bilayers(plus a Ga bilayer in the Ga-bilayer GaN(0001)-$\sqrt{3}\times \sqrt{3}$ system) to explore the preferred location of Co atoms in GaN surface.
In both two adopted configurations above, the upper three bilayers of GaN and adlayers
are allowed to relax, while the bottom bilayer of GaN and the saturating
pseudo-H atoms are fixed to mimic bulk substrate.
The vacuum region for the clean and Ga-bilayer surface are adopted as
$11 \mathrm{\AA}$ and $13 \mathrm{\AA}$ respectively.
Brillouin zones are sampled using $4\times 4\times 1$ and $6\times 6\times 1$
Monkhorst-Pack grids for the ($2\times 2$) and ($\sqrt{3}\times \sqrt{3}$) supercell,
respectively. All the atomic positions except the bottom GaN bilayers are relaxed until the
force exerted on each active atom was less than 0.03 eV/$\textrm{\AA}$.

\section{RESULTS AND DISCUSSIONS}

We first investigate various configurations (Ga-substitute, N-substitute and the interstitial site) with one Co atom doped in bulk GaN (including 72 atoms). The calculated formation energy for the $E_{Co_{Ga}}$ is of $2.94 eV$, which is relatively lower than that of $E_{Co_{N}}$(5.84 eV) and $E_{Co_{inter}}$(4.52 eV).
It can be concluded that Co atom prefers to locate at the Ga-substitutional site, which is in accordance with the experimental results from S. Dhara \cite{S. Dhara}.

\subsection{Energetics of Co-doped GaN(0001) surfaces}

\begin{figure}
    \begin{minipage}{\linewidth}
        \includegraphics[width=\textwidth]{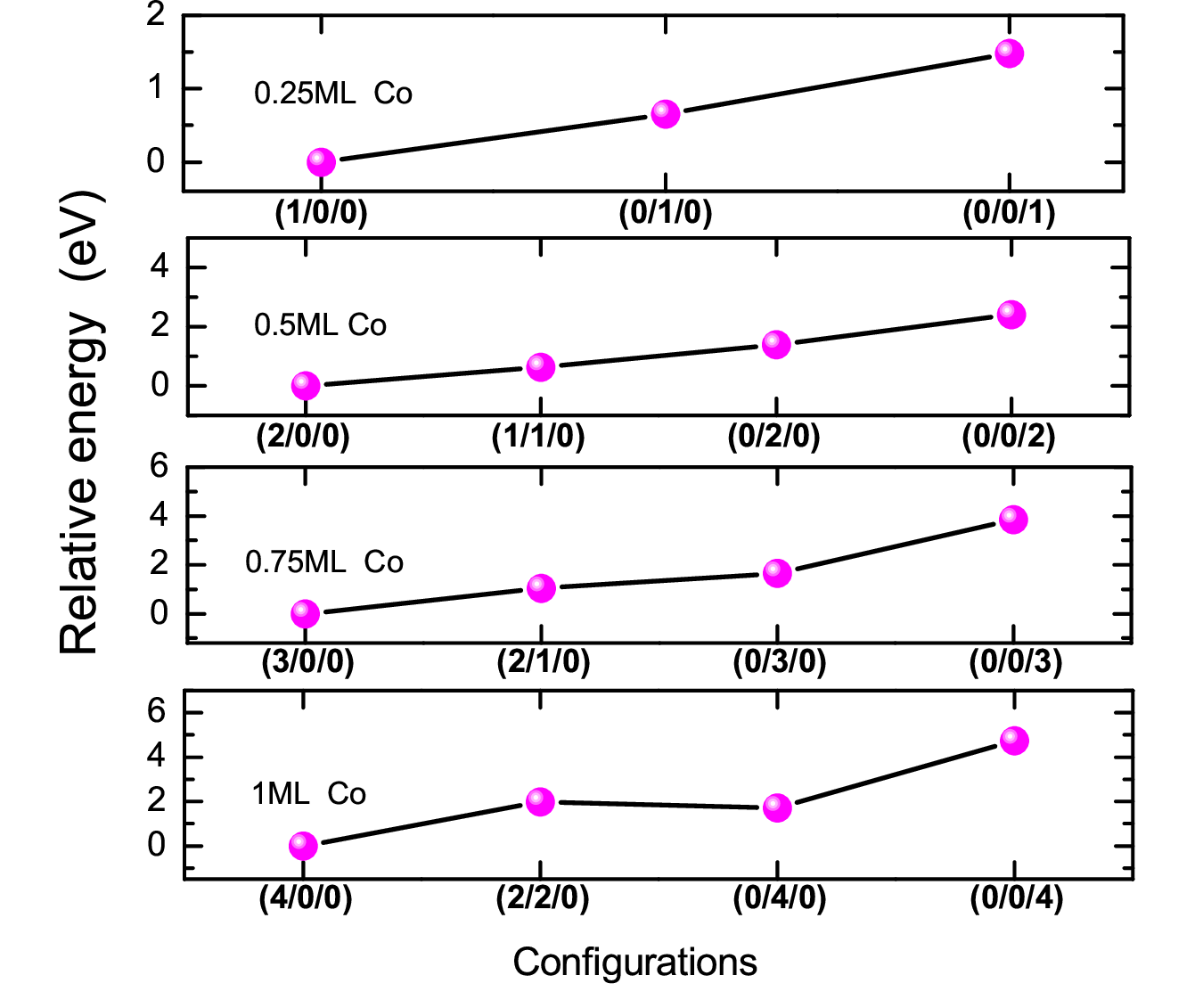}
    \end{minipage}
\caption{\label{Fig2}
(Color online)
Relative energies of different configurations for Co-doped clean
GaN(0001) surface.  Different configurations are labeled as $(n_1/n_2/n_3)$, where $n_1$, $n_2$ and $n_3$ are the Co atoms doped in the first, second, and third
GaN bilayer in a $2\times2$ supercell, respectively.}
\end{figure}

\begin{figure}
    \begin{minipage}{\linewidth}
        \includegraphics[width=\textwidth]{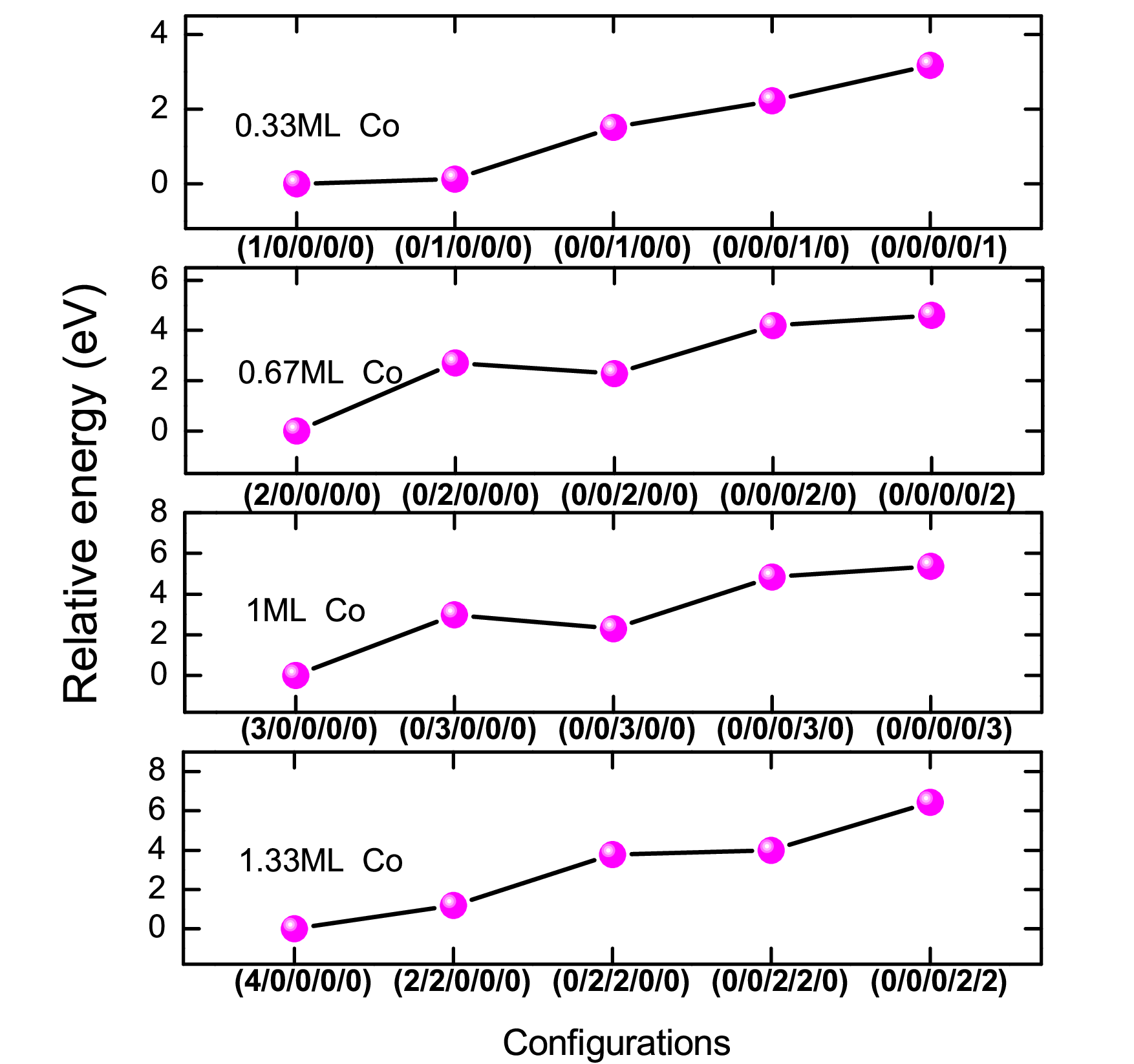}
    \end{minipage}
\caption{\label{Fig3}
(Color online)
Relative energies of different configurations for Co-doped Ga-bilayer GaN(0001) surface.  Different configurations are labeled as
$(n_1/n_2/n_3/n_4/n_5)$, where $n_1,n_2,n_3,n_4,n_5$ are the Co impurities incorporated in the
outer and inner Ga adlayers, first, second and third GaN bilayer in a $\sqrt{3}\times \sqrt{3}$
supercell, respectively.}
\end{figure}

For Co-doped clean GaN(0001) surface, the different configurations are denoted as $(n_1/n_2/n_3)$, where $n_1$, $n_2$ and $n_3$
are the numbers of substituting Co atoms in the first, second and third bilayer,
respectively(see Fig.~\ref{Fig1}(a)). Our results for the energetics of Co-doped clean GaN(0001) surface are summarized in
Fig.~\ref{Fig2}, including its Co-coverage dependence up to 1 monolayer(ML).

From Fig.~\ref{Fig2}, it appears that at 0.25 ML concentration, Co prefers to locate in the first bilayer. The total energy for Co in the second bilayer is about 0.65 $\mathrm{eV}$ higher, while the energy further increases by 1.48 $\mathrm{eV}$ for Co in the third bilayer.
As the Co concentration increased to 0.5 $\mathrm{ML}$, the (2/0/0) configuration becomes more stable
than the (1/1/0) with 0.62 $\mathrm{eV}$ lower in energy.
In addition, in the (2/0/0) configuration, the structure with Co atoms replacing two
closest Ga atoms is examined to be the most preferable.
As the Co concentration further increased to 0.75 $\mathrm{ML}$, the most stable
configuration still has all Co impurities in the first GaN bilayer, and the configurations (2/1/0) and (0/3/0) are 1.05 $\mathrm{eV}$ and 1.63 $\mathrm{eV}$ higher in energy, respectively.
At 1 $\mathrm{ML}$ Co concentration, the configuration
(4/0/0) becomes energetically stable.
Here we only consider the comparison with configurations of (2/2/0),
(0/4/0) and (0/4/0), which is already enough to find the distribution regularity.
Therefore, for Co-doped clean GaN(0001) surface, Co atoms prefer to
substitute the surface Ga sites.

On the other hand, site preference of Co-doped the Ga-bilayer GaN(0001) surface is investigated in the same way, $(n_1/n_2/n_3/n_4/n_5)$ is
used to denote the different incorporated configuration as shown in
Fig.~\ref{Fig1}(b). The relative energies for different configurations and concentrations of substitutional Co
impurities at the Ga-bilayer GaN(0001) surface are reported in Fig.~\ref{Fig3}, including its Co-coverage dependence up to 1.33 ML.

At 0.33 $\mathrm{ML}$ concentration, Co prefer to substitute the Ga atoms of outermost layer in Ga bilayer((1/0/0/0/0)), the total energy of this configuration is 1.51 $\mathrm{eV}$ lower than that of Co incorporated in the first GaN bilayer((0/0/1/0/0)). The optimized structure shows that Co atom and the underlying Ga atoms are connected tighter, which completely destroy the original flatting Ga bilayer structure.
As the Co concentration increased to 0.67 $\mathrm{ML}$, the additional impurity preferentially still occupies at Ga sites in the outermost layer of Ga bilayer((2/0/0/0/0)) and more Co atoms bond with the underlying Ga atoms.
Same situation continues to maintain as Co concentration further increases to 1 $\mathrm{ML}$, the
(3/0/0/0/0) configuration becomes the most stable.
As the Co concentration further increases to reach 1.33 $\mathrm{ML}$, the most
stable configuration is (4/0/0/0/0). Interestingly noted in the configuration(the optimized structure of (4/0/0/0/0) is shown in Fig. 7(b)), Co atoms replace the outermost Ga layer and connect closely with underlying Ga layer, which makes the original Ga bilayer structure turning to be a cluster-like CoGa bilayer.  According the calculation results, the average vertical distance between the newly formed Co layer and the underlying Ga layer is calculated as 1.93 $\textrm{\AA}$, which is significantly reduced compared to the original average distance(2.37 $\textrm{\AA}$) between the outermost Ga layer and underlying Ga layer. The surface characteristic and calculated data demonstrate that a flatting tightly coupled CoGa bilayer would be formed upon GaN bilayer in this case. Experimentally, CoGa alloys have been observed upon GaN(0001) substrate surface in system of Co deposition on GaN(0001) surface by MBE under excess Ga fluxes\cite{G. H. Zhong}.

Our results show that Co atoms tend to
concentrate in surface layer and do not show a migration to GaN bulk, which implies
that Co prefer to site at the surface, instead of incorporating in the bulk.  To summarize, the preferred site of Co-doped GaN(0001) surface
keeps always on the outermost surface, which is
completely different from our previous study on Al incorporation in GaN(0001)
surface that Al atoms prefer to migrate into deeper regions\cite{Zhenzhen Qin}.

With the relative energies calculated above, the stability of diagram for
Co:GaN(0001) can be derived as a function of Ga chemical potential to confirm
the growth mechanism.
The relative formation energy $E_{form}$ can be written as:
$$E_{form} = E_{total} - E_{refer} - \Delta n_{Co}\mu_{Co} - \Delta
n_{Ga}\mu_{Ga} - \Delta n_N \mu_N$$
Almost all relevant definitions were given in our previous study\cite{Zhenzhen Qin}. $\Delta
n_{Co}$ is the excess or deficit of Co atoms with respect to the reference.
A face centered cubic(fcc) crystal structure are adopted for Co-bulk.
The $\mu_{Co}$=$\mu_{Co(bulk)}$ corresponds to the Co-rich conditions.
 The obtained relative formation energy is a function of $\mu_{Ga}$ ranging from $\mu_{Ga(bulk)} + \Delta H_f^{GaN}$ to
$\mu_{Ga(bulk)}$.

\begin{figure}
    \begin{minipage}{\linewidth}
        \includegraphics[width=\textwidth]{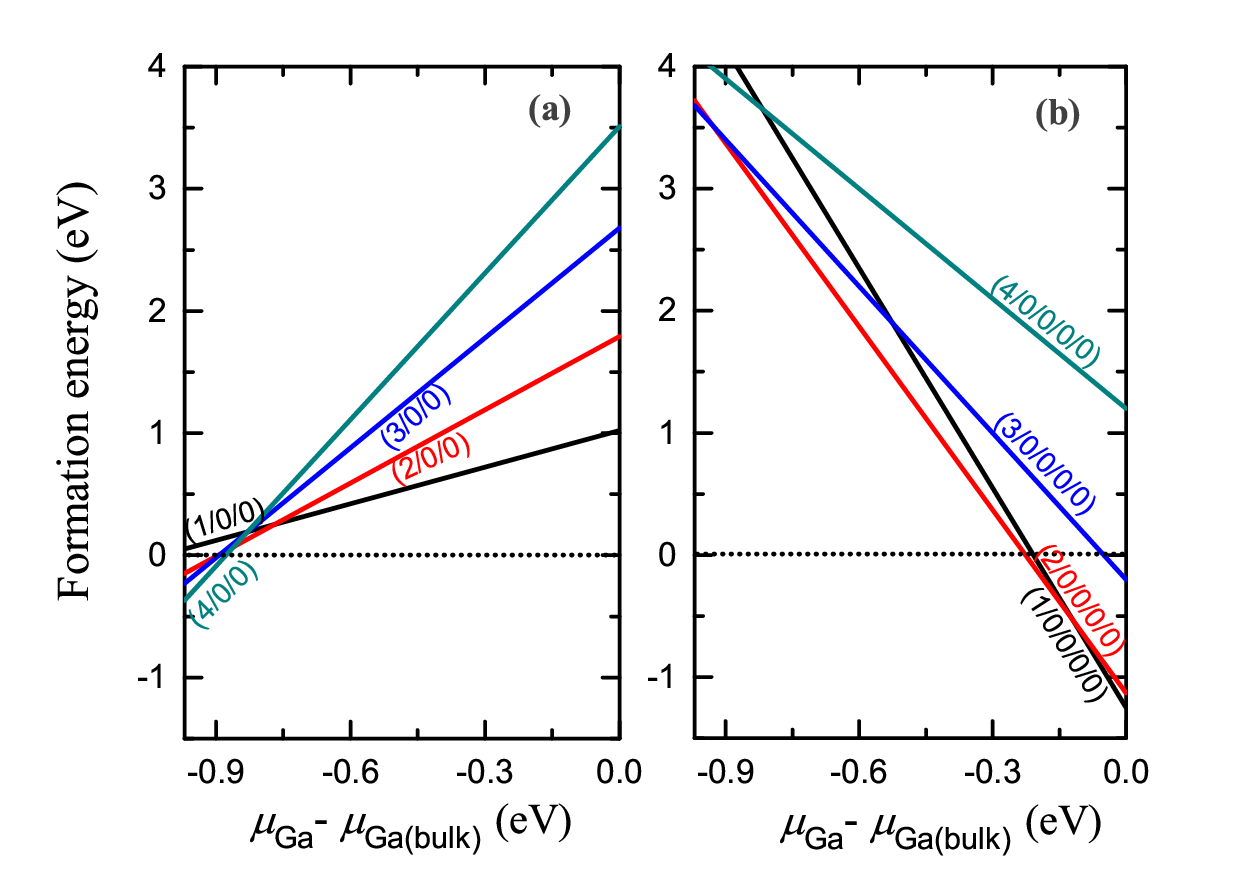}
    \end{minipage}
\caption{\label{Fig4}
(Color online)
Relative formation energies versus the Ga chemical potential $\mu_{Ga}$ for various configurations of (a) Co-doped clean GaN(0001) surfaces and (b) Co-doped Ga-bilayer GaN(0001) surfaces. The zero energy as indicated with a horizontal dot line corresponds to the total energies of $2\times 2$ and $\sqrt{3}\times \sqrt{3}$ supercell for the two surface respectively.}
\end{figure}

Fig.~\ref{Fig4} gives the the calculated formation energies for the most
favorable models of various Co concentration in two GaN(0001) surfaces. Under N-rich conditions, corresponding to the configurations of Co-doped clean GaN(0001) surface as shown in Fig.~\ref{Fig4}(a), it is found that the configuration with lowest energy is (4/0/0), meaning that a CoN/GaN interface structure could be formed with the CoN bilayer formed upon the GaN(0001) substrate surface.  Same interfacial behavior could also appeared in Fe doping GaN film\cite{iron}.
Furthermore, in the optimized structures of (4/0/0), the
bond length of Co-N is calculated as $1.90
\textrm{\AA}$, which is shorter than that of Ga-N as 1.97 $\textrm{\AA}$.  The doping of Co atoms does not destroy the original GaN fame with exhibiting a flatting appearance and encourages the bonding of Co atoms with N atoms, which caused the CoN bilayer matching well with GaN(0001) surface structure. Similar fundamental feature of CoZnO(0001) films growth and structures has been reported in previous experiment study\cite{A. Ney}.
While under the extremely Ga-rich conditions, corresponding to the configurations of Co-doped Ga-bilayer GaN(0001) surface as shown in Fig.~\ref{Fig4}(b), the Ga overlayer with an incorporated Co impurity becomes most favorable ((1/0/0/0/0)), indicating that CoGa$_x$/GaN interface is thermodynamically stable. According to our calculations, structures with higher Co concentration could be also energetically favorable except the configuration (4/0/0/0/0), the higher formation energy reveals that this surface structure would not be easily formed under natural growth environment, but might be achieved by changing external environment.

\subsection{Magnetic properties of Co-doped GaN(0001) surfaces}

\begin{figure}
    \begin{minipage}{\linewidth}
        \includegraphics[width=\textwidth]{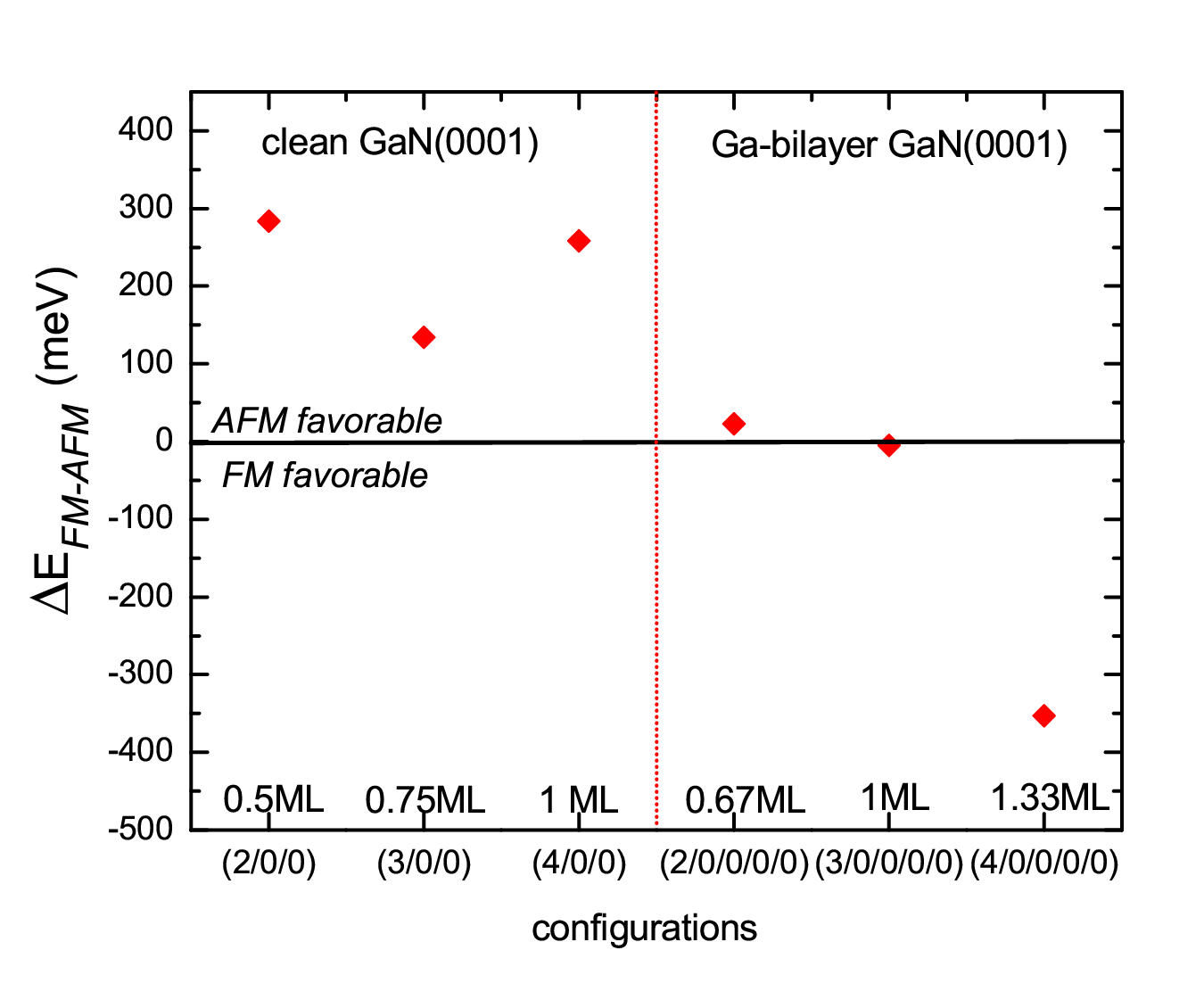}
    \end{minipage}
\caption{\label{Fig5}
(Color online)
Energy difference between the FM and AFM configurations as a function of various stable configurations in two cases of GaN(0001) surface. The zero energy is indicated with a horizontal line.}
\end{figure}

After achieving the most stable configuration for different Co concentrations, we investigate the magnetic coupling between the Co atoms in two GaN(0001) surfaces.
For Co-doped clean GaN(0001) surface, it is found that the magnetic moment of Co atom is 2.99 $\mu _{B}$ in the configuration (1/0/0), which is almost equal to a free Co atom situation (3 $\mu _{B}$).
Next, the stable (2/0/0), (3/0/0), (4/0/0) configurations are adopted to study the magnetism of Co-doped clean GaN(0001) surface in calculations. Both FM and AFM for each of these configurations are taken into consideration to identify the most preferred magnetic states. Interestingly, the energy differences between FM and AFM ($\Delta E_{FM-AFM}$) are calculated above zero, as shown in Fig.~\ref{Fig5}.  In detail, AFM is found to be more favorable over the FM by higher energy difference(284 meV) in the configuration (2/0/0), which indicates the AFM is relatively stable.  For the (3/0/0) and (4/0/0) configurations, the $\Delta E_{FM-AFM}$ are calculated as 134 and 259 meV respectively, indicating AFM state is still preferred over the FM state.  Therefore, it is concluded that the AFM coupling could be stable exist between Co atoms in clean GaN(0001) surface.  Our results provide an evidence for the previous experiment study which supposed the AFM state would appear in Ga$_{1-x}$Co$_{x}$N alloys under N-rich conditions\cite{S. Munawar Basha}. Coincidentally, same AFM phenomenon have been reported experimentally in CoZnO(0001) films\cite{Numan Akdogan, A. Ney}.

For Co-doped Ga-bilayer GaN(0001) surface, it is found that one Co atom could induce the magnetic moment of 1.12 $\mu _{B}$ into surface system. Then, the $\Delta E_{FM-AFM}$ of (2/0/0/0/0), (3/0/0/0/0), (4/0/0/0/0) configurations are calculated as shown in Fig.~\ref{Fig5}. Actually, the (2/0/0/0/0) configuration presents a marginal AFM behavior because of it's lower energy difference($\Delta E_{FM-AFM}$=23.2 meV).  But a tendency of FM behavior appears as Co concentration increased as shown in Fig.~\ref{Fig5}.  In detail, the (3/0/0/0/0) configuration shows FM state with a tiny energy difference ($\Delta E_{FM-AFM}$= -4.7 meV). What's interesting is that $\Delta E_{FM-AFM}$ of the (4/0/0/0/0) configuration shows a large negative value(-353 meV) with the total magnetic moment of 4.79 $\mu _{B}$. The calculated results demonstrate that FM coupling interaction will become more effective with increasing Co concentration in Ga-bilayer GaN(0001) surface, especially a relatively stable FM state is found at 1.33 $\mathrm{ML}$ concentration. But unfortunately, the higher formation energy of (4/0/0/0/0) configuration is a negative factor for obtaining such a stable FM structure.  It is therefore expected that the FM (4/0/0/0/0) configuration might be achieved based on some special experimental technologies in future, which may promise the potential applications of Co:GaN system as spin injection devices.
¡¡
\begin{figure}
    \begin{minipage}{\linewidth}
        \includegraphics[width=\textwidth]{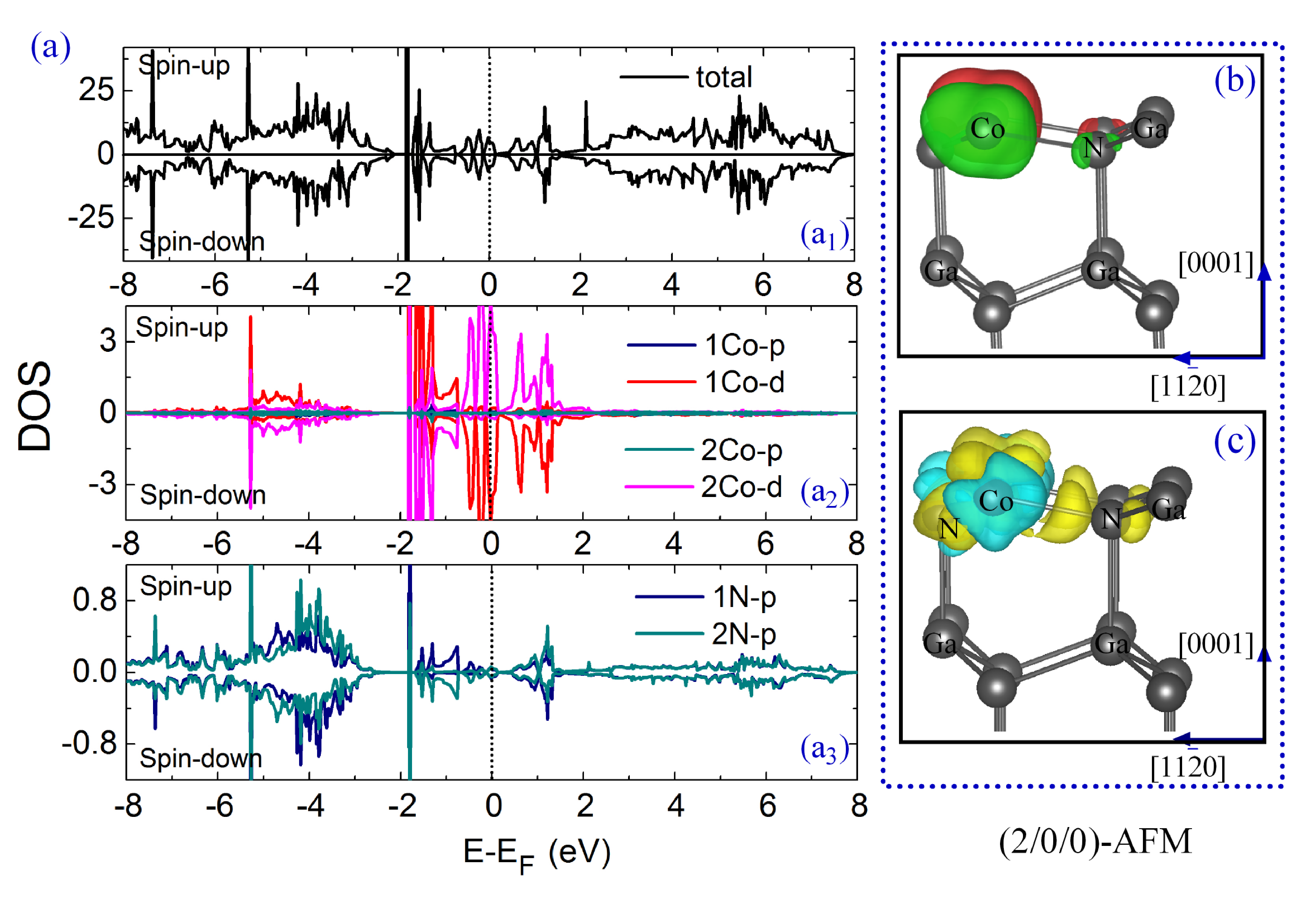}
    \end{minipage}
\caption{\label{Fig6}
(Color online) (a) DOS and partial DOS of the configuration (2/0/0). (b) Iso-surface plots(0.012 e/$\textrm{\AA}^3$) for spin density of configuration (2/0/0). Red color corresponds to the up-spin, and the green color corresponds to the down-spin. (c) Iso-surface plots(0.012 e/$\textrm{\AA}^3$) for charge density difference of the configuration (2/0/0).  Yellow contours indicate electron accumulation, and blue contours indicate electron depletion.}
\end{figure}

For the better understanding of the magnetic mechanism in Co-doped GaN(0001) surfaces, we perform the electronic structure calculations for both (2/0/0) and (4/0/0/0/0) configurations.

¡¡¡¡\begin{figure}
    \begin{minipage}{\linewidth}
        \includegraphics[width=\textwidth]{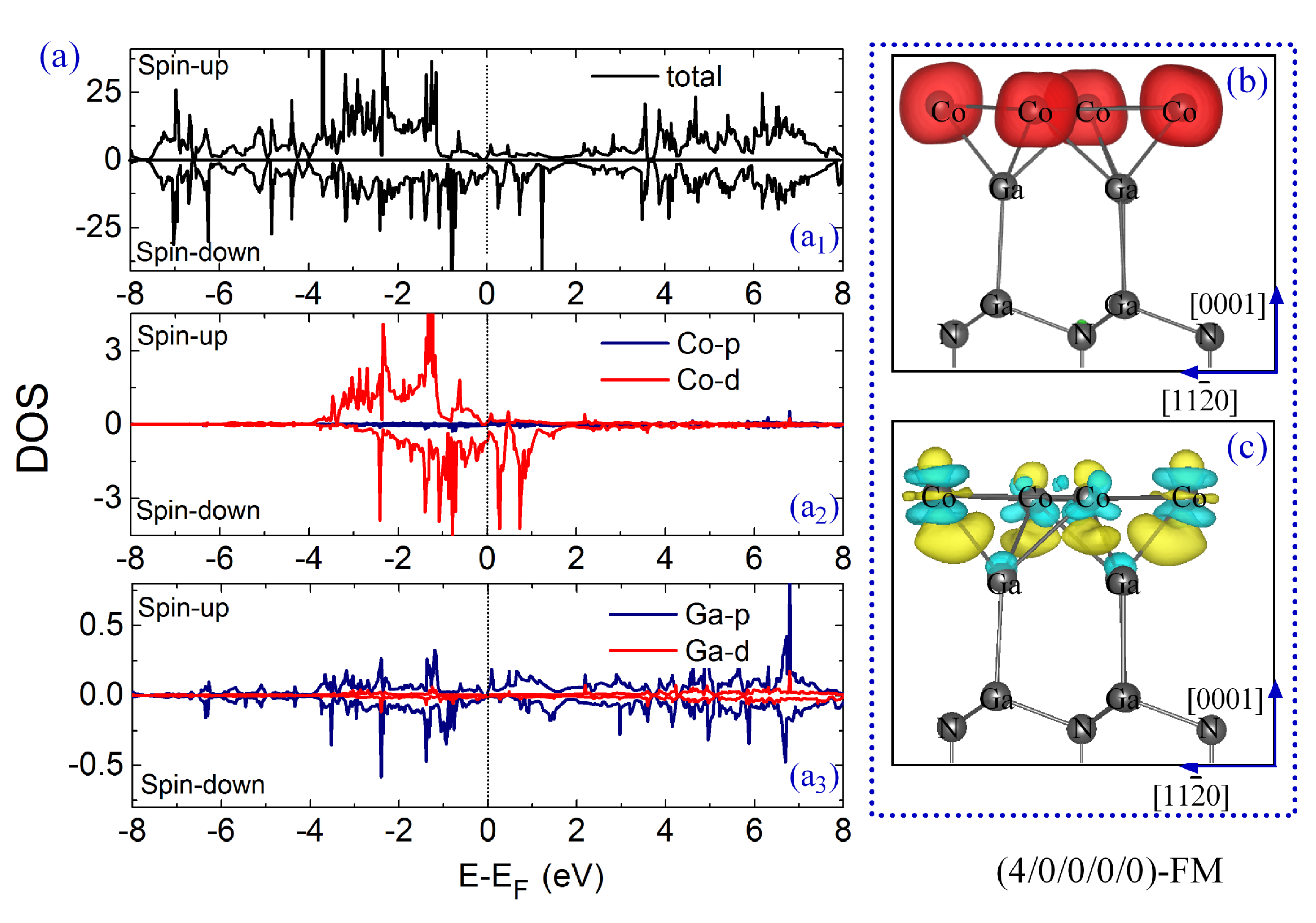}
    \end{minipage}
\caption{\label{Fig7}
(Color online) (a) DOS and partial DOS of the configuration (4/0/0/0/0). (b) Iso-surface plots(0.007 e/$\textrm{\AA}^3$) for spin-up and spin-down charge density of FM configuration (4/0/0/0/0). Red color corresponds to the up-spin, and the green color corresponds to the down-spin. (c)Iso-surface plots(0.007 e/$\textrm{\AA}^3$) for charge density difference of configuration (4/0/0/0/0).  Yellow contours indicate electron accumulation, and blue contours indicate electron depletion.}
\end{figure}¡¡
¡¡¡¡
The total density of states(DOS) for the (2/0/0) configuration, and the corresponding partial DOS of Co atoms and the partial DOS of neighboring N atoms are plotted in Fig. 6$(a_1)$, Fig. 6$(a_2)$, Fig. 6$(a_3)$, respectively. We note that the total DOS for spin-up and spin-down are identical, therefore there is no net magnetic moment in this case, as shown in Fig. 6$(a_1)$.  The magnetic moment on each one Co atom (refers to the marked 1Co or 2Co in Fig. 6$(a_2)$) is 1.84 $\mu _{B}$ and almost comes from the Co-3d orbital (1.79 $\mu _{B}$). Small contributions to the moment derive from the Co-4p (0.046 $\mu _{B}$) due to the p and d hybridization, see Fig. 6$(a_2)$. For each Co atom, the neighboring N atom is polarized non-reversing with a magnetic moment of 0.07 $\mu _{B}$ which mainly comes from N 2p orbital (0.06 $\mu _{B}$), as shown in Fig. 6$(a_3)$. From Fig. 6(b), it is clearly that two Co atoms and two neighboring N atoms are antiferromagnetically coupled respectively, and the total magnetic moment of the system is zero. We suppose that the antiferromagnetic ground state is due to the Co-N-Co superexchange interaction caused by the overlap of the Co-3d orbitals with the N-2p orbitals. Furthermore, the charge density difference of (2/0/0) is plotted in Fig. 6(c), which shows there is an obvious charge transfer from Co atom to neighboring N atom, which means a strong interaction between Co and N atoms.  This result offers an explanation for the previous experimental findings on the formation of Ga$_{1-x}$Co$_{x}$N in the Co-implanted GaN films under N-rich conditions\cite{G. H}.

The total DOS for spin-up and spin-down electrons states corresponding to the (4/0/0/0/0) configuration is plotted in Fig. 7$(a_1)$. The partial DOS of Co and neighboring Ga atoms are plotted in Fig. 7$(a_2)$ and Fig. 7$(a_3)$, respectively.
The total spin DOS shows that the spin-up and spin-down are not symmetric near $E_F$ with exhibited 100$\%$ polarization, therefore shows a half-metallic behavior and there is a net magnetic moment(4.79 $\mu _{B}$) in system.  The magnetic moment on each Co atom is 1.2 $\mu _{B}$ and mainly comes from Co-3d orbital (1.13 $\mu _{B}$), the remaining contribution to the magnetic moment arises from Co-4p, as shown in Fig. 7$(a_2)$. Actually, the electronic states of Co atoms and Ga atoms below contribute to the half-metallic behavior near the $E_F$ mainly due to the p-d(Ga-4p and Co-3d) hybridization, see Fig. 7$(a_2)$ and Fig. 7$(a_3)$, respectively.  Inversely, without the interaction of N atoms, the magnetic moments of four Co atoms are in the same parallel direction, leading to FM state in this case(see Fig. 7$(b)$). We suggest that the FM ground state mostly comes from the nearest-neighbor cation-cation coupling with the Co-Co direct exchange by the Co-3d orbitals, which is totally different with the configuration (2/0/0).
From Fig. 7$(c)$, we can see a delocalized enhancement of electron density at the Co-Ga bond, i.e., electron accumulation mostly localized between the outermost Co layer and underlying Ga layer, which indicates a strong interaction between Co and Ga atoms. Combining with the previous analysis of optimized geometric structure of (4/0/0/0/0) in Sec III.A, it is believed that this result could provides an intuitive understanding for the formation of CoGa alloys upon GaN(0001) substrate surface under excess Ga fluxes in experiment study\cite{G. H. Zhong}.

\section{CONCLUSIONS}
In summary, we have carried out a detail DFT study for the energetics and magnetic properties of two Co-doped GaN(0001) surfaces in different growth conditions. Our results suggest that substitution of surface Ga by Co is energetically favorable.
CoN/GaN interface could be easily formed under N-rich conditions, while under Ga-rich conditions, CoGa$_x$/GaN interface could be energetically favorable. Moreover, an anti-ferromagnetic ground state is favored for N-rich conditions, while the ground state is ferromagnetic for Ga-rich growth conditions.
We conclude that such magnetism in Co:GaN(0001) thin films is directly related to the growth conditions and Co:GaN system is expected to have potential applications in the field of spin injection devices.

\section{Acknowledgments}
The authors thank Prof. Xu Zuo of NKU for helpful discussions and acknowledge the computing
facilities of the Center for Computing of Tsinghua University. This work was supported by National Natural Science Foundation of China (Nos.51062003,61264005), Young Scientist Program of Jiangxi Province (20122BCB23030), Science Foundation of Jiangxi Province(Nos.GJJ14587, 20133BBE50001) and Innovation Team in JXSTNU (2013CXTD001).

\makeatletter
\renewcommand\@biblabel[1]{(#1)}


\begin{references}
\bibitem{H. Ohno} H. Ohno, Science, \textbf{281}, 5379, 951-956 (1998)
\bibitem{Shifei Qi} Shifei Qi, Fengxian Jiang, Jiuping Fan, H. Wu, S. B. Zhang, Gillian A. Gehring, Zhenyu Zhang, and Xiaohong Xu, Phys. Rev. B \textbf{84}, 205204 (2011)
\bibitem{Alexander A. Khajetoorians} Alexander A. Khajetoorians, Bruno Chilian,	Jens Wiebe,	 Sergej Schuwalow, Frank Lechermann and Roland Wiesendanger, Nature \textbf{467}, 1084¨C1087 (2010)	
\bibitem{Xu Zuo} Xu Zuo, Soack-Dae Yoon, Aria Yang, Wen-Hui Duan, Carmine Vittoria and Vincent G. Harris, J. Appl. Phys. \textbf{105}, 07C508 (2009)
\bibitem{S. Dhara} S. Dhara, B. Sundaravel, K. G. Nair, R. Kesavamoorthy, M. C. Valsakumar et al., Appl. Phys. Lett. \textbf{88}, 173110 (2006)
\bibitem{V. Baranwal} V. Baranwal, A. C. Pandey, J. W. Gerlach, B. Rauschenbach, H. Karl, D. Kanjilal and D. K. Avasthi, J. Appl. Phys. \textbf{103}, 124904 (2008)
\bibitem{G. H. Zhong} H. D. Li, G. H. Zhong, H. Q. Lin, and M. H. Xie, Phys. Rev. B \textbf{81}, 233302 (2010)
\bibitem{H D Li} H D Li, K He, M H Xie, N Wang, J F Jia and Q K Xue, New J. Phys. \textbf{12}, 073007 (2010)
\bibitem{W.Kim} Kim, W; Kang, H J; Oh, S K; Shin, S; Lee, J H; Song, J; Noh, S K; Oh, S J; Kim, CS, IEEE T. Nanotechnol. \textbf{5}, 2, 149 (2006)
\bibitem{S. Munawar Basha} S. Munawar Basha, S. Ramasubramanian, M. Rajagopalan, J. Kumar, Tae Won Kang,N. Ganapathi Subramaniam, Younghae Kwon, J Cryst. Growth, \textbf{318}, 432 (2011)
\bibitem{Yu-feng Tian} Yu-feng Tian, Yong-feng Li, and Tom Wu, Appl. Phys. Lett. 99, 222503 (2011)
\bibitem{W. G. Xie} W. G. Xie, F. Y. Xie, X. L. Yu, K. Xue, J. B. Xu, J. Chen and R. Zhang, Appl. Phys. Lett. 95, 262506 (2009)
\bibitem{A. Ney} A. Ney, K. Ollefs, S. Ye, T. Kammermeier, V. Ney, T. C. Kaspar, S. A. Chambers, F. Wilhelm, and A. Rogalev, Phys. Rev. Lett. 100, 157201 (2008)
\bibitem{Numan Akdogan} Numan Akdo$\breve{g}$an, Hartmut Zabel, Alexei Nefedov, Kurt Westerholt et.al,  J. Appl. Phys. 105, 043907 (2009)
\bibitem{G. Mula} G. Mula, C. Adelmann, S. Moehl, J. Oullier, B. Daudin, Phys. Rev. B \textbf{64}, 195406 (2001)
\bibitem{PhysRevB.69.035325} G. Koblm\"uller, R. Averbeck,  H. Riechert, P. Pongratz, Phys. Rev. B \textbf{69}, 035325 (2004)
\bibitem{J. E. Northrup} J. E. Northrup, J. Neugebauer, R. M. Feenstra, A. R. Smith, Phys. Rev. B \textbf{61}, 9932 (2000)
\bibitem{P.E.Blochl} P. E. Bl\"ochl, Phys. Rev. B 50, 17953(1994)
\bibitem{G.K} G. Kresse and J.Furthmuller, Comput. Mater. Sci. \textbf{6}, 15(1996)
\bibitem{J. P. Perdew} J. P. Perdew, K. Burke, M. Ernzerhof, Phys. Rev. Lett. \textbf{77}, 3865 (1996)
\bibitem{Zhenzhen Qin} Zhenzhen Qin, Zhihua Xiong, Guangzhao Qin, and Qixin Wan, J. Appl. Phys. \textbf{114}, 194307 (2013)
\bibitem{iron} Rafael Gonz$\acute{a}$lez-Hern$\acute{a}$ndez, William L$\acute{o}$pez P., Mar$\acute{\imath}$a G.Moreno-Armenta, Jairo Arbey Rodr$\acute{\imath}$guez, J. Appl. Phys. \textbf{109}, 07C102 (2011)
\bibitem{G. H} G. Husnain, Yao Shu-De, Ishaq Ahmad, H.M. Rafique , Arshad Mahmood, J. Magn. Magn. Mater. \textbf{324}, 797¨C801 (2012)






\end{references}
\end{document}